\documentclass[12pt]{iopart}

\usepackage{epsfig}
\usepackage{graphics}
\usepackage{graphicx}
\usepackage{subfigure}
\usepackage{multirow}

\begin{document}

\title[Particle Identification in the ALICE Experiment]{Particle Identification in the ALICE Experiment}

\author{Alexander Kalweit, for the ALICE collaboration}

\address{Instiut f\"ur Kernphysik, Technische Universit\"at Darmstadt and \\
Research Division and ExtreMe Matter Institute EMMI, \\
GSI Helmholtzzentrum f\"ur Schwerionenforschung, Darmstadt, Germany}
\ead{A.Kalweit@gsi.de}
\begin{abstract}

The particle identification capabilities of the ALICE experiment are unique among the four major LHC experiments. The working principles and excellent performance of the central barrel detectors in a high-multiplicity environment are presented as well as two physics examples: the extraction of transverse momentum spectra of charged pions, kaons, protons, and the observation of the ${\rm ^{4}\overline{He}}$-nucleus.

\end{abstract}

\vspace{-10mm}


\section{PID detectors and performance}

The ALICE experiment is exploiting almost all known techniques for particle identification (PID). The particle identification of the Inner Tracking System (ITS) and the Time Projection Chamber (TPC) are based on the specific energy loss per unit path length of a particle which depends for a given momentum only on its charge and rest mass. Thus, the simultaneous measurement of track momentum (or rigidity) and signal amplitude in a sensitive detector volume allows to identify particles. The measured mean energy deposit of a track is denoted as d$E$/d$x$ hereafter. In practice, it can be described with parameterizations of the well-known Bethe-Bloch formula.

Particle identification via d$E$/d$x$ needs to be supplemented by additional information  for momenta where the Bethe-Bloch curves for different particle species cross. The momenta of these tracks are high enough to reach the Time-Of-Flight detector (TOF) which is used to measure the particle's velocity and, with a momentum measurement, to determine its mass.  The reach of the hadron identification can be extended by the detection of Cherenkov radiation in the High Momentum Particle Identification Detector (HMPID) and a statistical analysis of the d$E$/d$x$ measurement on the relativistic rise in the TPC. High momenta electrons are identified via the detection of transition radiation. In the following section, the relevant PID detectors in the central barrel and their performance are described in increasing distance from the beam pipe.


\begin{figure}[htbp]
  \centering
  \includegraphics[width=1.\textwidth]{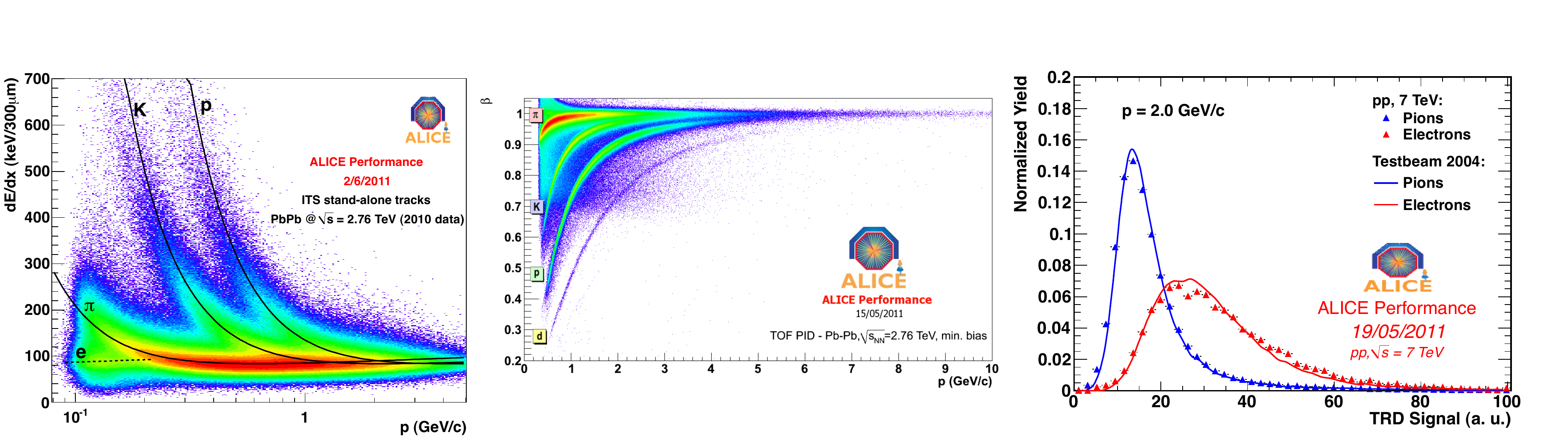}
  \caption{d$E$/d$x$ spectrum for the ITS (left), track velocity $\beta$ vs. momentum $p$ for the TOF (middle), and signal amplitude for electrons and pions in the TRD (right).}
  \label{figAllDetectorsITSTOF}
\end{figure}

%

\smallskip
{\bf Inner Tracking System (ITS).} Silicon Drift Detectors (SDD) and Silicon Strip Detectors (SSD), which form the two intermediate and two outer layers of ITS, respectively, provide analogue read-out for up to four samples for a truncated mean calculation of the d$E$/d$x$. Thus a resolution of $\sigma_{dE/dx} \approx 10-15\% $ is achieved (see fig.~\ref{figAllDetectorsITSTOF} left). The particle identification in the ITS combined with stand-alone tracking allows to identify pions with a minimum momentum of $p_{\rm t} \approx 100 \; \mathrm{MeV/}c$ which reduces the systematic error of yield and $\langle p_{\rm t}\rangle$  measurements due to extrapolation to $p_{\rm t} = 0$.

\smallskip
{\bf Time Projection Chamber (TPC).} The ALICE TPC with its 557$\,$568 readout channels \cite{Alme:2010ke} provides up to 159 ionization samples in a gas mixture of Ne and CO$_{2}$ (90\%/10\%). A truncated mean is used to reduce the Landau tail which results in a Gaussian distribution with a resolution of $\sigma_{dE/dx} \approx 5\% $. Figure \ref{figDeDx} (left) shows the measured d$E$/d$x$ versus the rigidity $R = p/z$ where $p$ corresponds to the track momentum and $z$ to the charge number. The lines show a parameterization of the Bethe-Bloch curve. The large dynamic range allows to detect particles with an average energy loss of 26 times minimum ionizing and thereby provides a clear identification of (anti-)nuclei.

\smallskip
{\bf Transition Radiation Detector (TRD).} Electron identification for momenta above 1 GeV/$c$ is achieved by the detection of transition radiation (see fig.~\ref{figAllDetectorsITSTOF} right). Transition radiation is produced by relativistic charged particles when they cross the many interfaces of two media of different dielectric constants in the radiator and is detected by the high-Z gas mixture of Xe and CO$_{2}$ (85\%/15\%).

\smallskip
{\bf Time-Of-Flight Detector (TOF).} The TOF detector is composed of 1638 multi-gap resistive plate chambers which provide an intrinsic resolution of approximately $80~{\rm ps}$. The overall time resolution for particle identification also depends on the time-0 uncertainty of the event. This results in a resolution of $\sigma_{TOF} = \sqrt{\sigma_{intr}^{2} + \sigma_{t0}^{2}} \approx 86~{\rm ps}$ for Pb--Pb collisions and $\sigma_{TOF} \approx 120~{\rm ps}$ for pp collisions. Hence a $2\sigma$-separation between protons and kaons up to 5 GeV/$c$ can be achieved in the high-multiplicity environment (see fig.~\ref{figAllDetectorsITSTOF} middle).

\smallskip
{\bf High Momentum Particle Identification Detector (HMPID).} The HMPID is a proximity focusing Ring Imaging Cherenkov detector with a liquid C$_{6}$F$_{14}$ radiator. It provides a separation of kaons and protons up to 5 GeV/$c$. 

\section{Single track and statistical particle identification}

In regions of clear separation, an identification of individual tracks is feasible, e.g. by assigning the particle type with the closest distance to an expected response function value. In practice, for detectors with a Gaussian response function this distance is usually specified in multiples of the resolution (so called n$\sigma$-cuts). These methods also find their application in {\it indirect} applications of particle identification, especially the removal of background in invariant mass analyses. The use of 3$\sigma$-cuts leads in many cases to a significant rejection of background without loss of efficiency. For the direct extraction of spectra, statistical unfolding methods can be applied in regions of only limited separation. The $p_{\rm t}$-reach of identified hadron spectra is further extended using the relativistic rise in the ionization measurement of the TPC.


	
	
	




\section{Topological particle identification}

\begin{figure}[htbp]
  \centering
  \subfigure{
    \label{figPionsPerDetector}
    \includegraphics[width=0.44\textwidth]{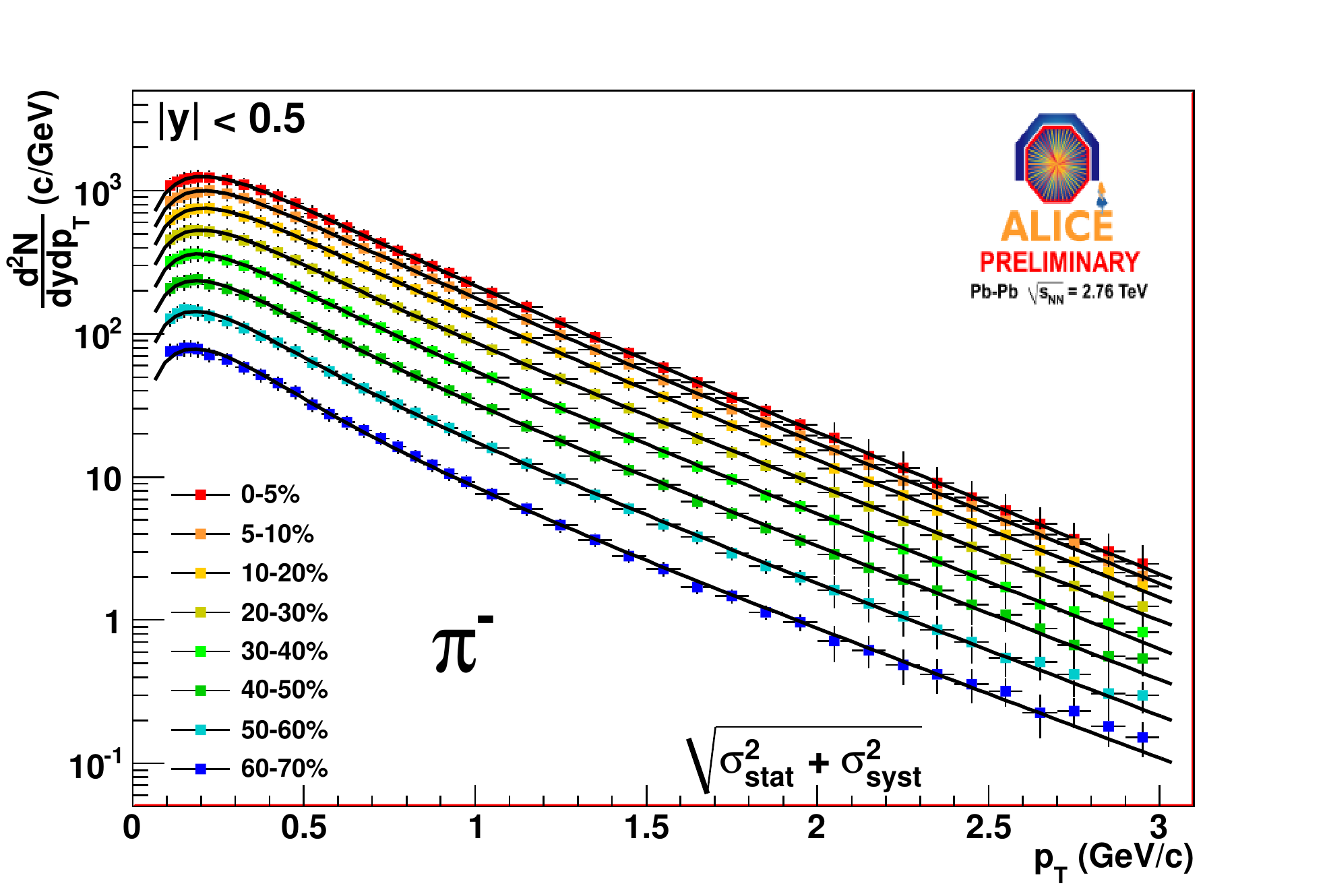} 
  }
  \subfigure{
    \label{figKaonComp}
    \includegraphics[width=0.44\textwidth]{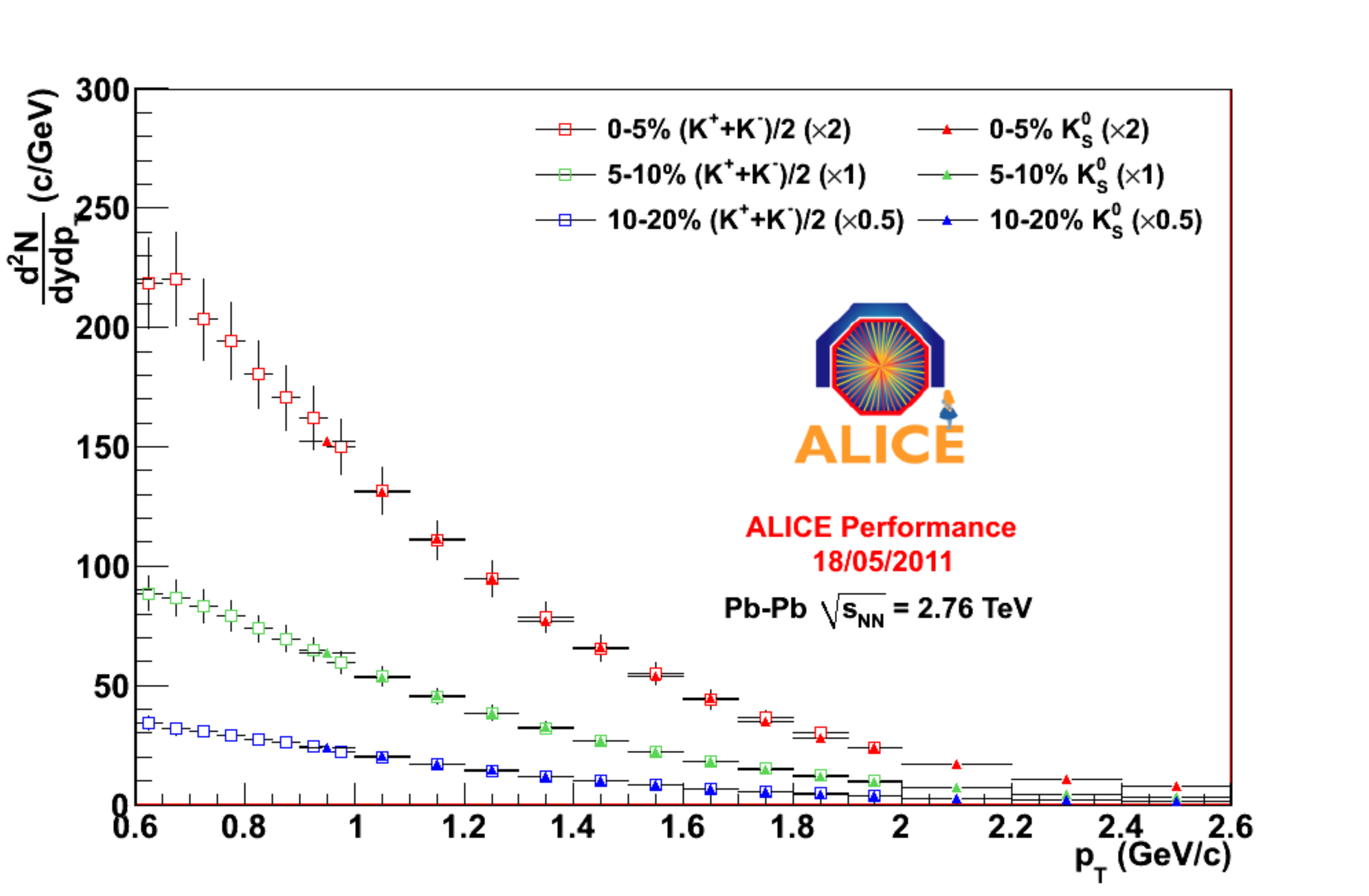} 
  }
  \caption{Combined pion spectra of the different analysis for several centralities (left) and comparison of charged and neutral kaons in Pb--Pb collisions (right).}
  \label{figSpectra}
\end{figure}

Weak decays of strange particles with a sufficiently long lifetime and $\gamma$-conversions can be identified via their characteristic decay topology \cite{Aamodt:2011zz}. This can be used to perform precise cross-checks between independent techniques. Spectra of kaons are obtained from five independent techniques: via the measurement of d$E$/d$x$, time-of-flight, Cherenkov radiation, as well as the V$^{0}$-type decay K$^{0}_{s} \rightarrow \pi^{+}\pi^{-}$, and the kink topology K$^{+} \rightarrow \mu^{+} \nu_{\mu}$. Figure \ref{figSpectra} illustrates the good agreement which is found for charged and neutral kaons.

\vspace{-10mm}
\section{Physics Example I: Spectra of $\pi^{\pm}$, ${\rm K^{\pm}}$, ${\rm p}$, and $\bar{{\rm p}}$}

The spectra of $\pi^{\pm}$, K$^{\pm}$, ${\rm p}$, and $\bar{{\rm p}}$ over a wide $p_{\rm t}$-range are extracted based on the methods outlined above \cite{Spectra900,ProceedingsMichele,ProceedingsMarek}. There, different strategies are followed by the individual detector projects: in the ITS and stand-alone TOF analysis, the distribution of the response function for tracks within $|y| <$ 0.5 is sliced in bins of p$_{t}$ and fitted with a superposition of Gaussian-like functions, to extract the yield of the different species following a statistical approach allowing a maximum $p_{\rm t}$-coverage of the individual detectors. The combined TPC-TOF analysis is based on a $3\sigma$-cut in the TPC at lower and in the TOF towards intermediate momenta where track-by-track PID is still possible. This allows to histogram the distance of closest approach to the primary vertex for each particle type. The raw yield of primary and secondary particles from weak decay or interaction with the detector material is then directly extracted via a fit of the corresponding Monte Carlo templates. Figure \ref{figSpectra} shows the combined pion spectrum of the different analyses. 


\section{Physics Example II: Observation of the ${\bf^{4}\overline{\bf{He}}}$-nucleus}

\begin{figure}[htbp]
  \centering
  \subfigure{
    \label{figUntriggeredDeDx}
    \includegraphics[width=0.53\textwidth]{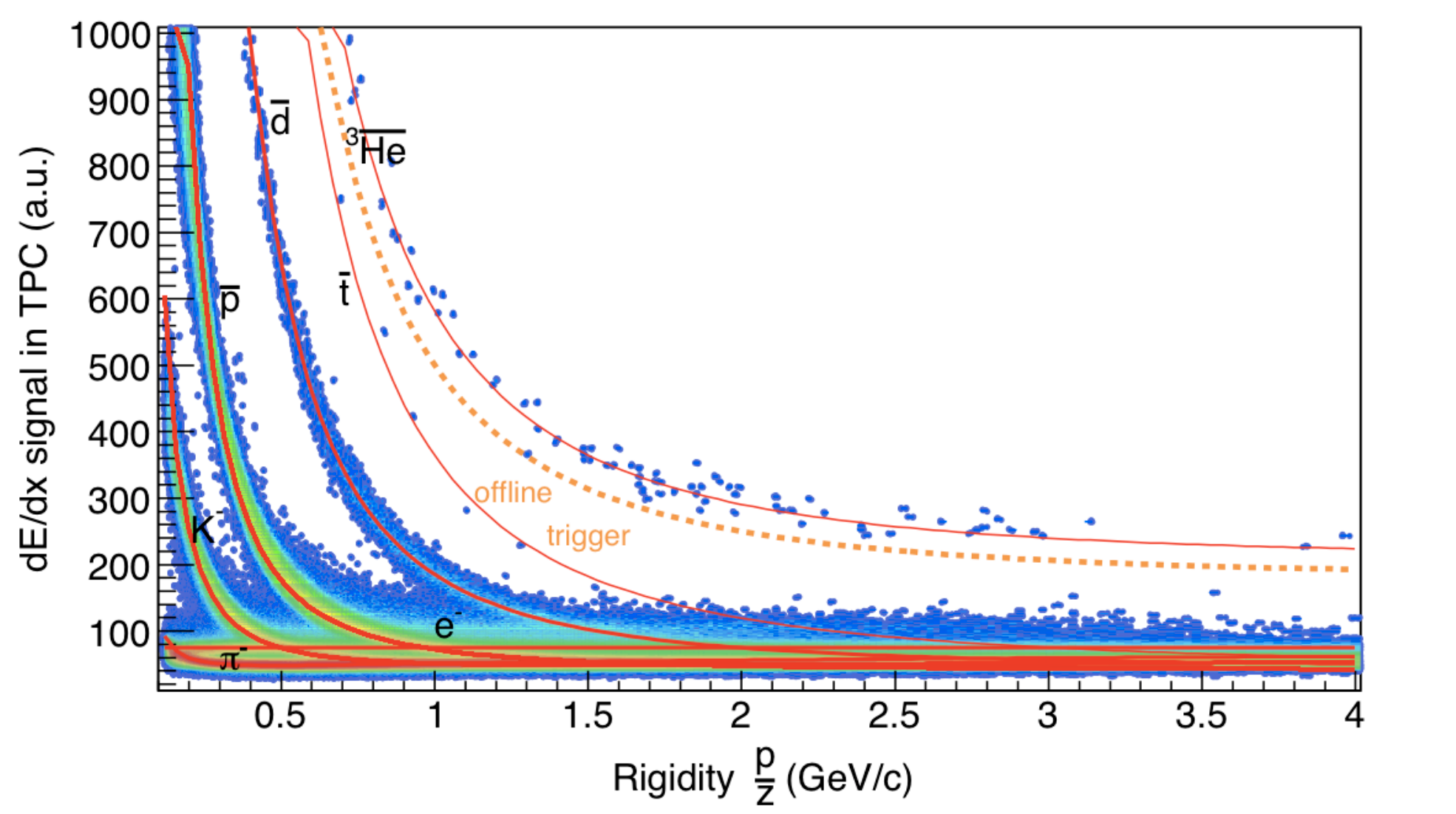} 
  }
  \subfigure{
    \label{figTriggeredDeDx}
    \includegraphics[width=0.4\textwidth]{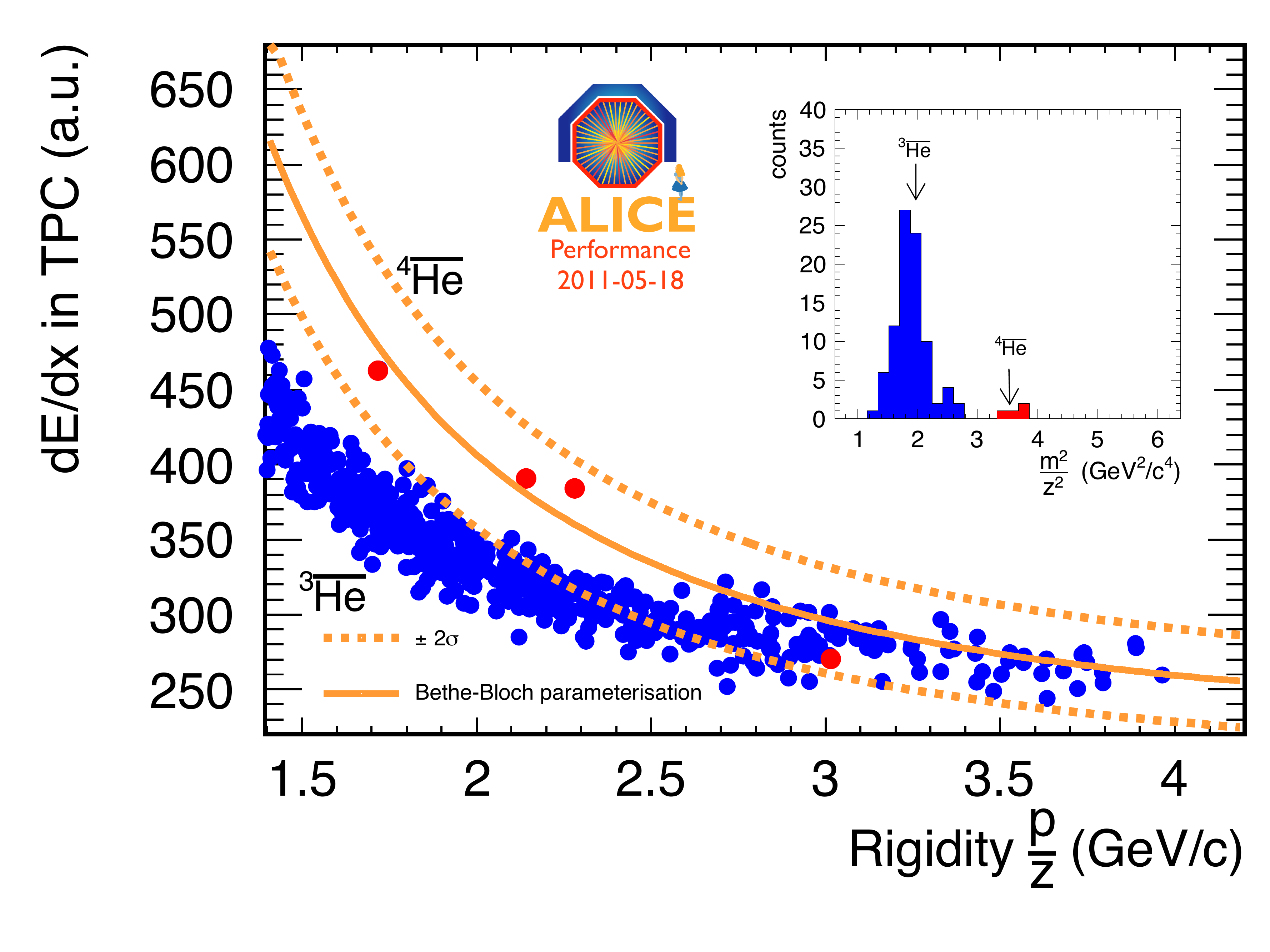} 
  }
  \caption{d$E$/d$x$ spectrum of the ALICE TPC of 2.2 million Pb--Pb events (left) and of the pre-selected tracks (by the offline trigger) in the full statistics (right). }
  \label{figDeDx}
\end{figure}

\noindent The ALICE experiment also observed four candidates of the ${^{4}\overline{\rm{He}}}$-nucleus, the measurement of which was recently published by the STAR collaboration \cite{Agakishiev:2011ib}. In total, 17.8 million nuclear collisions recorded in the heavy ion run of November 2010 were analyzed with an offline trigger selecting all ${^{3}\overline{\rm{He}}}$-nuclei or heavier. Figure \ref{figDeDx} (right) shows the d$E$/d$x$ versus rigidity distribution for negative particles in the region where the bands of ${^{3}\overline{\rm{He}}}$ and ${^{4}\overline{\rm{He}}}$ are clearly visible. Below a rigidity of $p/z \approx 2.2 \; {\rm GeV}/c$ two candidates are clearly identified only based on the d$E$/d$x$ information. Above, the mass determination of the candidate tracks must be combined with mass determined with the TOF system following 

\begin{equation}
 \label{eqMassTOF}
 m^{2} / z^{2} = R^{2} / (\gamma^{2} - 1)  \; .
\end{equation}

\noindent The inlet in figure \ref{figDeDx} (right) shows the ${m^{2} \over z^{2}}$ distribution for all tracks within a 2$\sigma$-band around the expected d$E$/d$x$ for ${^{4}\overline{\rm{He}}}$. The four anti-alpha candidates are highlighted in red in both the ${m^{2} \over z^{2}}$ and the d$E$/d$x$ versus rigidity plot. The d$E$/d$x$ cut selects particles such that only tracks with $z = 2$ are contained in the sample which removes the ambiguity with deuterons (see eq. \ref{eqMassTOF}).

\end{document}